\documentclass[11pt]{article}
\usepackage{graphicx,amsmath}
\usepackage{hyperref}

\begin{document}
%\begin{document}
%\special{papersize=8.26in,11.69in}
%textwidth17.0cm
%\textheight22.0cm
%\baselineskip1.0cm
%\setlength{\topmargin}{-1cm}
%\addtolength{\textheight}{1cm}
%\oddsidemargin+0.8cm
%\evensidemargin-0.8cm
\pagestyle{plain}
\def\eqa{\!\!&=&\!\!}
\def\ccr{\nonumber\\}

\def\la{\langle}
\def\ra{\rangle}

\def\del{\Delta}
\def\ddel{{}^\bullet\! \Delta}
\def\deld{\Delta^{\hskip -.5mm \bullet}}
\def\ddeld{{}^{\bullet}\! \Delta^{\hskip -.5mm \bullet}}
\def\dddel{{}^{\bullet \bullet} \! \Delta}

\def\rld{\rlap{\,/}D}
\def\rldd{\rlap{\,/}\nabla}
%%%%%%%%%%%%%%% insert actual file mydefs.sty %%%%%%%%%%%%%%%%%%%%%%
%------------------------------------------------------------------------     
% MATH SYMBOLS
%
%fractions
\def\half{{1\over 2}}
\def\third{{1\over3}}
\def\fourth{{1\over4}}
\def\fifth{{1\over5}}
\def\sixth{{1\over6}}
\def\seventh{{1\over7}}
\def\eigth{{1\over8}}
\def\ninth{{1\over9}}
\def\tenth{{1\over10}}
\def\bN{\mathop{\bf N}}
\def\R{{\rm I\!R}}
\def\Eins{{\mathchoice {\rm 1\mskip-4mu l} {\rm 1\mskip-4mu l}
{\rm 1\mskip-4.5mu l} {\rm 1\mskip-5mu l}}}
\def\Z{{\mathchoice {\hbox{$\sf\textstyle Z\kern-0.4em Z$}}
{\hbox{$\sf\textstyle Z\kern-0.4em Z$}}
{\hbox{$\sf\scriptstyle Z\kern-0.3em Z$}}
{\hbox{$\sf\scriptscriptstyle Z\kern-0.2em Z$}}}}
\def\abs#1{\left| #1\right|}
\def\com#1#2{
        \left[#1, #2\right]}
\def\square{\kern1pt\vbox{\hrule height 1.2pt\hbox{\vrule width 1.2pt
   \hskip 3pt\vbox{\vskip 6pt}\hskip 3pt\vrule width 0.6pt}
   \hrule height 0.6pt}\kern1pt}
      \def\boxop{{\raise-.25ex\hbox{\square}}}
% \contract is a differential geometry contraction sign _|
\def\contract{\makebox[1.2em][c]{
        \mbox{\rule{.6em}{.01truein}\rule{.01truein}{.6em}}}}
\def\ltap{\ \raisebox{-.4ex}{\rlap{$\sim$}} \raisebox{.4ex}{$<$}\ }
\def\gtap{\ \raisebox{-.4ex}{\rlap{$\sim$}} \raisebox{.4ex}{$>$}\ }
\def\mn{{\mu\nu}}
\def\rs{{\rho\sigma}}
\newcommand{\Det}{{\rm Det}}
\def\Tr{{\rm Tr}\,}
\def\tr{{\rm tr}\,}
\def\sumij{\sum_{i<j}}
\def\e{\,{\rm e}}
%derivatives
\def\pa{\partial}
\def\dA{\partial^2}
\def\ddx{{d\over dx}}
\def\ddt{{d\over dt}}
\def\der#1#2{{d #1\over d#2}}
\def\lie{\hbox{\it \$}} % fancy L for the Lie derivative
\def\partder#1#2{{\partial #1\over\partial #2}}
\def\secder#1#2#3{{\partial^2 #1\over\partial #2 \partial #3}}
%
%equations
\newcommand{\be}{\begin{equation}}
\newcommand{\ee}{\end{equation}\noindent}
\newcommand{\bear}{\begin{eqnarray}}
\newcommand{\ear}{\end{eqnarray}\noindent}
\newcommand{\benn}{\begin{enumerate}}
\newcommand{\enn}{\end{enumerate}}
\newcommand{\veject}{\vfill\eject}
\newcommand{\ven}{\vfill\eject\noindent}
%
%reference to equations
\def\eq#1{{eq. (\ref{#1})}}
\def\eqs#1#2{{eqs. (\ref{#1}) -- (\ref{#2})}}
%
%integrals
\def\totint{\int_{-\infty}^{\infty}}
\def\posint{\int_0^{\infty}}
\def\negint{\int_{-\infty}^0}
\def\pint{{\dps\int}{dp_i\over {(2\pi)}^d}}
%
% PHYS SYMBOLS
\newcommand{\GeV}{\mbox{GeV}}
\def\FFdual{F\cdot\tilde F}
\def\bra#1{\langle #1 |}
\def\ket#1{| #1 \rangle}
\def\braket#1#2{\langle {#1} \mid {#2} \rangle}
\def\vev#1{\langle #1 \rangle}
\def\rightvac{\mid 0\rangle}
\def\leftvac{\langle 0\mid}
\def\ihbar{{i\over\hbar}}
% dirac matrix stuff
\def\ge{\hbox{$\gamma_1$}}
\def\gz{\hbox{$\gamma_2$}}
\def\gd{\hbox{$\gamma_3$}}
\def\go{\hbox{$\gamma_1$}}
\def\gt{\hbox{\$\gamma_2$}}
\def\gth{\hbox{$\gamma_3$}} 
\def\gf{\hbox{$\gamma_5\;$}}
\newcommand{\slD}{\,\raise.15ex\hbox{$/$}\kern-.27em\hbox{$\!\!\!D$}}
\newcommand{\slpartial}{\raise.15ex\hbox{$/$}\kern-.57em\hbox{$\partial$}}
\newcommand{\cL}{\cal L}
\newcommand{\D}{\cal D}
\newcommand{\Dhalf}{{D\over 2}}
\def\eps{\epsilon}
\def\epshalf{{\epsilon\over 2}}
\def\lag{( -\partial^2 + V)}
%worldline
\def\freeexp{{\rm e}^{-\int_0^Td\tau {1\over 4}\dot x^2}}
\def\kinb{{1\over 4}\dot x^2}
\def\kinf{{1\over 2}\psi\dot\psi}
\def\expk{{\rm exp}\biggl[\,\sum_{i<j=1}^4 G_{Bij}k_i\cdot k_j\biggr]}
\def\expp{{\rm exp}\biggl[\,\sum_{i<j=1}^4 G_{Bij}p_i\cdot p_j\biggr]}
\def\expshort{{\e}^{\half G_{Bij}k_i\cdot k_j}}
\def\expabb{{\e}^{(\cdot )}}
\def\epseps#1#2{\varepsilon_{#1}\cdot \varepsilon_{#2}}
\def\epsk#1#2{\varepsilon_{#1}\cdot k_{#2}}
\def\kk#1#2{k_{#1}\cdot k_{#2}}
\def\G#1#2{G_{B#1#2}}
\def\Gp#1#2{{\dot G_{B#1#2}}}
\def\GF#1#2{G_{F#1#2}}
\def\Dab{{(x_a-x_b)}}
\def\Dsq{{({(x_a-x_b)}^2)}}
\def\PITD{{(4\pi T)}^{-{D\over 2}}}
\def\4piTD{{(4\pi T)}^{-{D\over 2}}}
\def\4piT4{{(4\pi T)}^{-2}}
\def\TintmD{{\dps\int_{0}^{\infty}}{dT\over T}\,e^{-m^2T}
    {(4\pi T)}^{-{D\over 2}}}
\def\Tintm4{{\dps\int_{0}^{\infty}}{dT\over T}\,e^{-m^2T}
    {(4\pi T)}^{-2}}
\def\Tintm{{\dps\int_{0}^{\infty}}{dT\over T}\,e^{-m^2T}}
\def\Tint{{\dps\int_{0}^{\infty}}{dT\over T}}
\def\np{n_{+}}
\def\nm{n_{-}}
\def\Np{N_{+}}
\def\Nm{N_{-}}
\newcommand{\slG}{{{\dot G}\!\!\!\! \raise.15ex\hbox {/}}}
\newcommand{\Gd}{{\dot G}}
\newcommand{\Gund}{{\underline{\dot G}}}
\newcommand{\Gdd}{{\ddot G}}
\def\GBd12{{\dot G}_{B12}}
\def\Dx{\dps\int{\cal D}x}
\def\Dy{\dps\int{\cal D}y}
\def\Dpsi{\dps\int{\cal D}\psi}
\def\dint#1{\int\!\!\!\!\!\int\limits_{\!\!#1}}
\def\ddtau{{d\over d\tau}}
\def\ie{\hbox{$\textstyle{\int_1}$}}
\def\iz{\hbox{$\textstyle{\int_2}$}}
\def\id{\hbox{$\textstyle{\int_3}$}}
\def\ldop{\hbox{$\lbrace\mskip -4.5mu\mid$}}
\def\rdop{\hbox{$\mid\mskip -4.3mu\rbrace$}}
%
%VARIOUS
\newcommand{\1}{{\'\i}}
\newcommand{\no}{\noindent}
\def\non{\nonumber}
\def\dps{\displaystyle}
\def\sy{\scriptscriptstyle}
\def\sy{\scriptscriptstyle}
%-------------------------------------------------------------------------

%-------------------------------------------------------------------------
%% My own definitions
% From cv.sty
%
% Title page

\newcommand{\bea}{\begin{eqnarray}}  
\newcommand{\eea}{\end{eqnarray}}  
\def\eqa{&=&}  
\def\ccr{\nonumber\\}  
  
\def\a{\alpha}
\def\b{\beta}
\def\m{\mu}
\def\n{\nu}
\def\r{\rho}
\def\s{\sigma}
\def\ep{\epsilon}

\def\cosech{\rm cosech}
\def\sech{\rm sech}
\def\coth{\rm coth}
\def\tanh{\rm tanh}

%%%%%%%%  
\def\sqr#1#2{{\vcenter{\vbox{\hrule height.#2pt  
     \hbox{\vrule width.#2pt height#1pt \kern#1pt  
           \vrule width.#2pt}  
       \hrule height.#2pt}}}}  
\def\square{\mathchoice\sqr66\sqr66\sqr{2.1}3\sqr{1.5}3}  
%%%%%%%%%%%  
  
\def\appendix{\par\clearpage
  \setcounter{section}{0}
  \setcounter{subsection}{0}
 % \@addtoreset{equation}{section}
  \def\@sectname{Appendix~}
  \def\theequation{\thesection\arabic{equation}}
  \def\thesection{\Alph{section}}}
 
% Figures
\def\thefigures#1{\par\clearpage\section*{Figures\@mkboth
  {FIGURES}{FIGURES}}\list
  {Fig.~\arabic{enumi}.}{\labelwidth\parindent\advance
\labelwidth -\labelsep
      \leftmargin\parindent\usecounter{enumi}}}
\def\figitem#1{\item\label{#1}}
\let\endthefigures=\endlist
 
% Tables
\def\thetables#1{\par\clearpage\section*{Tables\@mkboth
  {TABLES}{TABLES}}\list
  {Table~\Roman{enumi}.}{\labelwidth-\labelsep
      \leftmargin0pt\usecounter{enumi}}}
\def\tableitem#1{\item\label{#1}}
\let\endthetables=\endlist
 
% Put period after section number and allow for APPENDIX prefix.
\def\@sect#1#2#3#4#5#6[#7]#8{\ifnum #2>\c@secnumdepth
     \def\@svsec{}\else
     \refstepcounter{#1}\edef\@svsec{\@sectname\csname the#1\endcsname
.\hskip 1em }\fi
     \@tempskipa #5\relax
      \ifdim \@tempskipa>\z@
        \begingroup #6\relax
          \@hangfrom{\hskip #3\relax\@svsec}{\interlinepenalty \@M #8\par}
        \endgroup
       \csname #1mark\endcsname{#7}\addcontentsline
         {toc}{#1}{\ifnum #2>\c@secnumdepth \else
                      \protect\numberline{\csname the#1\endcsname}\fi
                    #7}\else
        \def\@svse=chd{#6\hskip #3\@svsec #8\csname #1mark\endcsname
                      {#7}\addcontentsline
                           {toc}{#1}{\ifnum #2>\c@secnumdepth \else
                             \protect\numberline{\csname the#1\endcsname}\fi
                       #7}}\fi
     \@xsect{#5}}
 
\def\@sectname{}
%
%                 M A T E X
%
%       This defines et al., i.e., e.g., cf., etc.
\def\eg{\hbox{\it e.g.}}        \def\cf{\hbox{\it cf.}}
\def\etal{\hbox{\it et al.}}
\def\dash{\hbox{---}}
%       common physics symbols
\def\bR{\mathop{\bf R}}
\def\bC{\mathop{\bf C}}
\def\eq#1{{eq. \ref{#1}}}
\def\eqs#1#2{{eqs. \ref{#1}--\ref{#2}}}
\def\lie{\hbox{\it \$}} % fancy L for the Lie derivative
\def\partder#1#2{{\partial #1\over\partial #2}}
\def\secder#1#2#3{{\partial^2 #1\over\partial #2 \partial #3}}
\def\abs#1{\left| #1\right|}
\def\ltap{\ \raisebox{-.4ex}{\rlap{$\sim$}} \raisebox{.4ex}{$<$}\ }
\def\gtap{\ \raisebox{-.4ex}{\rlap{$\sim$}} \raisebox{.4ex}{$>$}\ }
% \contract is a differential geometry contraction sign _|
\def\contract{\makebox[1.2em][c]{
        \mbox{\rule{.6em}{.01truein}\rule{.01truein}{.6em}}}}
% double-headed superior arrow added 9.2.86
%
% commutator added 11.14.86
\def\com#1#2{
        \left[#1, #2\right]}
%
%these written by orlando alvarez
% ************************************************************
%       The following macros were written by Chris Quigg.
%       They create bent arrows and can be used to write
%       decays such as pi --> mu + nu
%                              --> e nu nubar
%
\def\bentarrow{\:\raisebox{1.3ex}{\rlap{$\vert$}}\!\rightarrow}
\def\longbent{\:\raisebox{3.5ex}{\rlap{$\vert$}}\raisebox{1.3ex}%
        {\rlap{$\vert$}}\!\rightarrow}
\def\onedk#1#2{
        \begin{equation}
        \begin{array}{l}
         #1 \\
         \bentarrow #2
        \end{array}
        \end{equation}
                }
\def\dk#1#2#3{
        \begin{equation}
        \begin{array}{r c l}
        #1 & \rightarrow & #2 \\
         & & \bentarrow #3
        \end{array}
        \end{equation}
                }
\def\dkp#1#2#3#4{
        \begin{equation}
        \begin{array}{r c l}
        #1 & \rightarrow & #2#3 \\
         & & \phantom{\; #2}\bentarrow #4
        \end{array}
        \end{equation}
                }
\def\bothdk#1#2#3#4#5{
        \begin{equation}
        \begin{array}{r c l}
        #1 & \rightarrow & #2#3 \\
         & & \:\raisebox{1.3ex}{\rlap{$\vert$}}\raisebox{-0.5ex}{$\vert$}%
        \phantom{#2}\!\bentarrow #4 \\
         & & \bentarrow #5
        \end{array}
        \end{equation}
                }

\begin{center}
{\Large\bf On the low energy limit of one loop photon-graviton amplitudes}\\

\bigskip

{F. Bastianelli$^{a,b}$, O. Corradini$^{c}$, J.~M. D\'avila$^{d}$ and 
C. Schubert$^{b,d}$}
\begin{itemize}
\item [$^a$]
{\it
Dipartimento di Fisica, Universit\`a di Bologna
  and\\ INFN, Sezione di Bologna, Via Irnerio 46, I-40126
  Bologna, Italy
  }
  \item[$^b$]
  {\it 
%   \footnote{}
Max-Planck-Institut f\"ur Gravitationsphysik, Albert-Einstein-Institut,
M\"uhlenberg 1, D-14476 Potsdam, Germany
}
\item [$^c$] 
 {\it 
%   \footnote{}
Centro de Estudios en F\'isica y Matem\'aticas B\'asicas y Aplicadas\\
Universidad Aut\'onoma de Chiapas\\ C.P. 29000, Tuxtla Guti\'errez, M\'exico
}
\item [$^d$]
{\it 
Instituto de F\'{\i}sica y Matem\'aticas
\\
Universidad Michoacana de San Nicol\'as de Hidalgo\\
Edificio C-3, Apdo. Postal 2-82\\
C.P. 58040, Morelia, Michoac\'an, M\'exico\\
}
\end{itemize}
\end{center}

\bigskip

\noindent
{\bf Abstract:}
We present first results of a systematic study of the structure of the low energy limit of the one-loop photon-graviton amplitudes induced by massive scalars and spinors. 
Our main objective is the search of KLT-type relations where effectively two photons merge into a graviton.
We find such a relation at the graviton-photon-photon level.
We also derive the diffeomorphism Ward identity for the 1PI one graviton - $N$ photon amplitudes.

\vfill\eject\noindent

\renewcommand{\theequation}{\arabic{section}.\arabic{equation}}
\renewcommand{\arraystretch}{2.5}
\def\R{1\!\!{\rm R}}
\def\Eins{\mathord{1\hskip -1.5pt
\vrule width .5pt height 7.75pt depth -.2pt \hskip -1.2pt
\vrule width 2.5pt height .3pt depth -.05pt \hskip 1.5pt}}
\newcommand{\symb}{\mbox{symb}}
\renewcommand{\arraystretch}{2.5}
\def\GBd12{{\dot G}_{B12}}
\def\mneg{\!\!\!\!\!\!\!\!\!\!}
\def\Mneg{\!\!\!\!\!\!\!\!\!\!\!\!\!\!\!\!\!\!\!\!}
\def\non{\nonumber}
\def\beqn*{\begin{eqnarray*}}
\def\eqn*{\end{eqnarray*}}
\def\sy{\scriptscriptstyle}
\def\footstrut{\baselineskip 12pt}
\def\square{\kern1pt\vbox{\hrule height 1.2pt\hbox{\vrule width 1.2pt
   \hskip 3pt\vbox{\vskip 6pt}\hskip 3pt\vrule width 0.6pt}
   \hrule height 0.6pt}\kern1pt}
\def\np{n_{+}}
\def\nm{n_{-}}
\def\Np{N_{+}}
\def\Nm{N_{-}}
\def\exmn{\Bigl(\mu \leftrightarrow \nu \Bigr)}
\def\slash#1{#1\!\!\!\!\raise.15ex\hbox {\hspace{3pt}/}}
\def\dint#1{\int\!\!\!\!\!\int\limits_{\!\!#1}}
\def\bra#1{\langle #1 |}
\def\ket#1{| #1 \rangle}
\def\vev#1{\langle #1 \rangle}
\def\rightvac{\mid 0\rangle}
\def\leftvac{\langle 0\mid}
\def\dps{\displaystyle}
\def\sy{\scriptscriptstyle}
\def\half{{1\over 2}}
\def\third{{1\over3}}
\def\fourth{{1\over4}}
\def\fifth{{1\over5}}
\def\sixth{{1\over6}}
\def\seventh{{1\over7}}
\def\eigth{{1\over8}}
\def\ninth{{1\over9}}
\def\tenth{{1\over10}}
\def\pa{\partial}
\def\ddtau{{d\over d\tau}}
\def\ge{\hbox{\textfont1=\tame $\gamma_1$}}
\def\gz{\hbox{\textfont1=\tame $\gamma_2$}}
\def\gd{\hbox{\textfont1=\tame $\gamma_3$}}
\def\go{\hbox{\textfont1=\tamt $\gamma_1$}}
\def\gt{\hbox{\textfont1=\tamt $\gamma_2$}}
\def\gth{\hbox{\textfont1=\tamt $\gamma_3$}} 
\def\gf{\hbox{$\gamma_5\;$}}
\def\ie{\hbox{$\textstyle{\int_1}$}}
\def\iz{\hbox{$\textstyle{\int_2}$}}
\def\id{\hbox{$\textstyle{\int_3}$}}
\def\ldop{\hbox{$\lbrace\mskip -4.5mu\mid$}}
\def\rdop{\hbox{$\mid\mskip -4.3mu\rbrace$}}
\def\eps{\epsilon}
\def\epshalf{{\epsilon\over 2}}
\def\e{\mbox{e}}
\def\mn{{\mu\nu}}
\def\exmn{{(\mu\leftrightarrow\nu )}}
\def\ab{{\alpha\beta}}
\def\exab{{(\alpha\leftrightarrow\beta )}}
\def\g{\mbox{g}}
\def\kinb{{1\over 4}\dot x^2}
\def\kinf{{1\over 2}\psi\dot\psi}
\def\expk{{\rm exp}\biggl[\,\sum_{i<j=1}^4 G_{Bij}k_i\cdot k_j\biggr]}
\def\expp{{\rm exp}\biggl[\,\sum_{i<j=1}^4 G_{Bij}p_i\cdot p_j\biggr]}
\def\expshort{{\e}^{\half G_{Bij}k_i\cdot k_j}}
\def\expabb{{\e}^{(\cdot )}}
\def\epseps#1#2{\varepsilon_{#1}\cdot \varepsilon_{#2}}
\def\epsk#1#2{\varepsilon_{#1}\cdot k_{#2}}
\def\kk#1#2{k_{#1}\cdot k_{#2}}
\def\G#1#2{G_{B#1#2}}
\def\Gp#1#2{{\dot G_{B#1#2}}}
\def\GF#1#2{G_{F#1#2}}
\def\Dab{{(x_a-x_b)}}
\def\Dsq{{({(x_a-x_b)}^2)}}
\def\lag{( -\partial^2 + V)}
\def\PITD{{(4\pi T)}^{-{D\over 2}}}
\def\4piTD{{(4\pi T)}^{-{D\over 2}}}
\def\4piT4{{(4\pi T)}^{-2}}
\def\TintmD{{\dps\int_{0}^{\infty}}{dT\over T}\,e^{-m^2T}
    {(4\pi T)}^{-{D\over 2}}}
\def\Tintm4{{\dps\int_{0}^{\infty}}{dT\over T}\,e^{-m^2T}
    {(4\pi T)}^{-2}}
\def\Tintm{{\dps\int_{0}^{\infty}}{dT\over T}\,e^{-m^2T}}
\def\Tint{{\dps\int_{0}^{\infty}}{dT\over T}}
\def\pint{{\dps\int}{dp_i\over {(2\pi)}^d}}
\def\Dx{\dps\int{\cal D}x}
\def\Dy{\dps\int{\cal D}y}
\def\Dpsi{\dps\int{\cal D}\psi}
\def\Tr{{\rm Tr}\,}
\def\tr{{\rm tr}\,}
\def\sumij{\sum_{i<j}}
\def\freeexp{{\rm e}^{-\int_0^Td\tau {1\over 4}\dot x^2}}
\def\arraystretch{2.5}
\def\Ge{\mbox{GeV}}
\def\dA{\partial^2}
\def\DA{\sqsubset\!\!\!\!\sqsupset}
\def\FFdual{F\cdot\tilde F}
\def\mn{{\mu\nu}}
\def\rs{{\rho\sigma}}
\def\oplusotimes{{{\lower 15pt\hbox{$\scriptscriptstyle \oplus$}}\atop{\otimes}}}
\def\perppar{{{\lower 15pt\hbox{$\scriptscriptstyle \perp$}}\atop{\parallel}}}
\def\oopp{{{\lower 15pt\hbox{$\scriptscriptstyle \oplus$}}\atop{\otimes}}\!{{\lower 15pt\hbox{$\scriptscriptstyle \perp$}}\atop{\parallel}}}
%\font\tame = cmmi12 scaled\magstep1
%\font\tamt = cmmi12 scaled\magstep2
%-------------------------------------------------------------------------
% To change the LaTeX pagestyle
% example  DINA4 format DESY
%\newlength{\dinwidth}
%\newlength{\dinmargin}
%\setlength{\dinwidth}{21.0cm}
%\textheight23.2cm
%\textwidth17.0cm
%\setlength{\dinmargin}{\dinwidth}
%\addtolength{\dinmargin}{-\textwidth}
%\setlength{\dinmargin}{0.5\dinmargin}
%\oddsidemargin -1.0in
%\addtolength{\oddsidemargin}{\dinmargin}
%\setlength{\evensidemargin}{\oddsidemargin}
%\setlength{\marginparwidth}{0.9\dinmargin}
%\marginparsep 8pt \marginparpush 5pt
%\topmargin -42pt
%\headheight 12pt 
%\headsep 30pt \footheight 12pt \footskip
%24pt
%-----------------------------------------------------------------------
% uncomment any of these if you want numbering to respect the sections
%
% \renewcommand{\thesection}{\arabic{section}.}
% \renewcommand{\thesubsection}{\thesection\arabic{subsection}.}
% \renewcommand{\theequation}{{\protect\thesection\arabic{equation}}}
% \renewcommand{\thetable}{{\protect{\bf \thesection\arabic{table}}}}
% \renewcommand{\thetable}{{\protect{\thesection\arabic{table}}}}
% \renewcommand{\thefigure}{{\protect\bf\thesection\arabic{figure}}}
% \renewcommand{\thefigure}{{\protect\thesection\arabic{figure}}}
% \renewcommand{\textfraction}{0}
% \renewcommand{\topfraction}{1.00}
% \renewcommand{\bottomfraction}{1.00}
% \renewcommand{\baselinestretch}{1.1}
%-----------------------------------------------------------------------
% special symbols: real numbers, unit matrix, integers
%
\def\bbbr{{\rm I\!R}}
\def\bbbone{{\mathchoice {\rm 1\mskip-4mu l} {\rm 1\mskip-4mu l}
{\rm 1\mskip-4.5mu l} {\rm 1\mskip-5mu l}}}
\def\bbbz{{\mathchoice {\hbox{$\sf\textstyle Z\kern-0.4em Z$}}
{\hbox{$\sf\textstyle Z\kern-0.4em Z$}}
{\hbox{$\sf\scriptstyle Z\kern-0.3em Z$}}
{\hbox{$\sf\scriptscriptstyle Z\kern-0.2em Z$}}}}

%%%%%%%%%%%%%%%%%%%%%%%%%%%%%%%%%%%%%%%%%%%%%%%%%%%%%%%%%%%%%%%%%%%%%%%
\renewcommand{\thefootnote}{\protect\arabic{footnote}}
%\pagestyle{plain}
%------------------------------------------------------
%\hfill {\large AEI-2008-053}
%
%\hfill {\large UMSNH-IFM-F-2008-24}

%\pagestyle{plain}
%\setcounter{page}{1}
%\setcounter{footnote}{0}

%\topmargin=-.3in
%\textheight=8.3in
%\textwidth=7.2in
%\textwidth=6.7in

\vspace{10pt}
\section{Introduction}
\renewcommand{\theequation}{1.\arabic{equation}}
\setcounter{equation}{0}

Although graviton amplitudes are presently not of phenomenological relevance, 
their structure has been studied in parallel with the more important Yang-Mills amplitudes.
The powerful methods that have been developed during the last two decades
for the computation of on-shell amplitudes 
(see, e.g., \cite{bedush,bddk,bedika,brcafe,bbst,cacsvr,beboca,bcdjr}) essentially apply equally to both cases.
Moreover, the famous relations found in 1986 by  Kawai, Lewellen and Tye (KLT) between closed and open string amplitudes
\cite{kalety} in the field theory limit imply relations between amplitudes in gravity and
Yang-Mills theory that are not at all obvious from the field theory Lagrangians or Feynman rules \cite{begiku,bdpr,bern}. 
The effort to make these relations transparent also at the field theory level is still ongoing 
\cite{bergra,bencac,anathe,bdfs,dhsz}.
The KLT relations express gravity amplitudes as sums of squares of Yang-Mills amplitudes.
They hold at the tree level, but can be used together with unitarity methods for 
 the construction of loop amplitudes in gravity. 
This is very interesting considering the different UV behaviour of loop amplitudes in gravity vs. Yang-Mills theory,
and has been used as a tool in the study of the possible finiteness of $N=8$ supergravity (see \cite{bcdjr,fermar} and refs. therein).

More recently, a different kind of KLT-like relations has been found 
where the same type of factorization is made manifest even at the integrand level.
Reversing the original flow of information,
these relations were first conjectured for the $n$ - graviton tree amplitudes in field theory \cite{becajo:2008},
and later extended to and proven in string theory \cite{bjdava,stieberger}.
A multiloop generalization in field theory has been conjectured in \cite{becajo:2010}. 

Most of this work concerned the case of massless on-shell amplitudes,
for which particularly efficient computation methods are available.
Much less has been done on amplitudes involving the
interaction of gravitons with massive matter. At the tree level, there are some
classical results on amplitudes involving gravitons \cite{bergas,milton}.
More recently, the tree-level Compton-type amplitudes involving gravitons and
spin zero, half and one particles were computed in \cite{holstein}, leading to another remarkable
factorization property \cite{chshso} of the graviton-graviton scattering amplitudes in terms
of the photonic Compton amplitudes. 

There are also a number of results for mixed graviton-gauge boson amplitudes
involving a matter loop, namely
the graviton-photon-photon vertex \cite{druhat,bafrts,shore2002,giamot,arcodePRD81}, 
its nonabelian generalization \cite{arcodePRD82}, and the related amplitude for
photon-graviton conversion in an external field \cite{gertsenshtein,rafsto,phograv1,phograv2,ahjari}. 

Here we will present first results of a systematic study of the structure of the mixed
photon-graviton amplitudes with a massive loop in the low energy limit, and of the
search for KLT-like relations for such amplitudes. 
The great advantage of this limit is that it is accessible through the effective action;
let us discuss this first for the purely photonic case.
As is well-known, the information on  the low energy limit of the QED $N$ photon amplitudes
is contained in the Euler-Heisenberg Lagrangian (``EHL'') \cite{eulhei}. We recall the standard representation
of this effective Lagrangian:

\bear
{\cal L}^{(EH)}_{\rm spin} &=& - {1\over 8\pi^2}
\int_0^{\infty}{dT\over T^3}
\,\e^{-m^2T}
\biggl\lbrack
{(eaT)(ebT)\over {\rm tanh}(eaT)\tan(ebT)} 
%\nonumber\\&&\hspace{70pt}
- {e^2\over 3}(a^2-b^2)T^2 -1
\biggr\rbrack \, .
\non\\
\label{eulhei}
\ear
Here $T$ is the proper-time of the loop fermion, $m$ its mass, and $a,b$ are the two
Maxwell field invariants,
related to $\bf E$, $\bf B$ by $a^2-b^2 = B^2-E^2,\quad ab = {\bf E}\cdot {\bf B}$.
The subtraction terms implement the renormalization of charge and vacuum
energy.
The analogous representation for scalar QED was obtained by Weisskopf \cite{weisskopf}:

\begin{eqnarray}
{\cal L}^{(EH)}_{\rm scal}(F)&=&  {1\over 16\pi^2}
\int_0^{\infty}{dT\over T^3}
\,{\rm e}^{-m^2T} 
\biggl[
{(eaT)(ebT)\over \sinh(eaT)\sin(ebT)} 
%\nonumber\\&&\hspace{110pt}
+{1\over 6}(a^2-b^2)T^2 -1
\biggr] \non\\
\label{weisskopf}
\end{eqnarray}
Obtaining the low energy ( = large mass) limit of the $N$ photon amplitudes
from the effective Lagrangians (\ref{eulhei}), (\ref{weisskopf}) is a standard procedure
(see, e.g., \cite{itzzub-book}),  and the result can be expressed quite concisely \cite{mascvi}:

\bear
A_{\rm spin}
[\varepsilon_1^+;\ldots ;\varepsilon_K^+;\varepsilon_{K+1}^-;\ldots ;\varepsilon_N^-]
&=&
-{m^4\over 8\pi^2}
\Bigl({2ie\over m^2}\Bigr)^N(N-3)!
\non\\&&\hspace{-90pt}\times
\sum_{k=0}^K\sum_{l=0}^{N-K}
(-1)^{N-K-l}
{{\cal B}_{k+l}{\cal B}_{N-k-l}
\over
k!l!(K-k)!(N-K-l)!}
\chi_K^+\chi_{N-K}^- 
\, ,
\non\\
A_{\rm scal}
[\varepsilon_1^+;\ldots ;\varepsilon_K^+;\varepsilon_{K+1}^-;\ldots ;\varepsilon_N^-]
&=&
{m^4\over 16\pi^2}
\Bigl({2ie\over m^2}\Bigr)^N(N-3)!
\non\\&&\hspace{-130pt}\times
\sum_{k=0}^K\sum_{l=0}^{N-K}
(-1)^{N-K-l}
{\bigl(1-2^{1-k-l)}\bigr)
\bigl(1-2^{1-N+k+l}\bigr)
B_{k+l}B_{N-k-l}
\over
k!l!(K-k)!(N-K-l)!}
\chi_K^+\chi_{N-K}^-  
\, .
\non\\
\label{EHamp}
\ear
Here the superscripts $\pm$ refer to circular polarizations, and the ${\cal B}_k$ are
Bernoulli numbers. The invariants $\chi_K^{\pm}$ are written, in spinor helicity
notation (our spinor helicity conventions follow \cite{dixonrev}),

\bear
\chi_K^+ &=&
{({\frac{K}{2}})!
\over 2^{K\over 2}}
\Bigl\lbrace
[12]^2[34]^2\cdots [(K-1)K]^2 + {\rm \,\, all \,\, permutations}
\Bigr\rbrace,
\non\\
\chi_K^- &=&
{({\frac{K}{2}})!
\over 2^{K\over 2}}
\Bigl\lbrace
\langle 12\rangle^2\langle3 4\rangle^2\cdots \langle (K-1)K\rangle^2 + {\rm \,\, all \,\, permutations}
\Bigr\rbrace.
\non\\
\label{defchi}
\ear 
For the case of the ``maximally helicity-violating'' (MHV) amplitudes, which have
``all $+$'' or ``all $-$'' helicities, eqs. (\ref{EHamp}) simplify (using Bernoulli number
identities) to 

\bear
A_{\rm scal}
[k_1,\varepsilon_1^{\pm};\ldots ;k_N,\varepsilon_N^{\pm}]
&=&-
\frac{(2e)^{N}}{(4\pi)^2m^{2N-4}}
\frac{{\cal B}_{N}}{N(N-2)}
\chi_N^{\pm} 
\label{ampscalSD}\\
A_{\rm spin}
[k_1,\varepsilon_1^{\pm};\ldots ;k_N,\varepsilon_N^{\pm}]
&=&-2
A_{\rm scal}
[k_1,\varepsilon_1^{\pm};\ldots ;k_N,\varepsilon_N^{\pm}]
\label{relscalspin}
\ear
This relation (\ref{relscalspin}) is actually true also away from the low-energy limit, 
and can be explained by the fact that the MHV amplitudes correspond to a 
self-dual background, in which the Dirac operator has
a quantum-mechanical supersymmetry \cite{dufish}.
For this MHV case eqs. (\ref{EHamp}) have also
been generalized to the two-loop level \cite{dunschsd1}. 

One of the long-term goals of the line of work presented here is to obtain a generalization of
(\ref{EHamp}) to the case of the mixed $N$ - photon / $M$ - graviton amplitudes.
As a first step, in \cite{badasc} the EHL (\ref{eulhei}) and its scalar analogue
were generalized to the effective actions corresponding to the low energy 
one graviton - $N$ photon amplitudes. Those were obtained in \cite{badasc} 
in terms of two-parameter integrals (a similar result was found in \cite{avrfuc}).  
Expanding out the spinor loop Lagrangian in powers of field invariants 
one finds, up to total deriative terms, the  
Lagrangian obtained in the seminal work of  Drummond and Hathrell \cite{druhat},

\bear
{\cal L}_{\rm spin}^{(DH)} &=& 
\frac{e^2}{180 (4\pi)^2m^2} 
\bigg(
5 R F_{\mu\nu}^2 
-26 R_{\mu\nu} F^{\mu\alpha} F^\nu{}_\alpha
+2  R_{\mu\nu\alpha\beta}F^{\mu\nu}F^{\alpha\beta} \non\\
&& \qquad\qquad
+24 (\nabla^\alpha F_{\alpha\mu})^2  
\bigg )
\label{drumhath}
\ear
(see \cite{badasc} for our gravity conventions). 
The corresponding form of the effective Lagrangian for the scalar loop
(using the same operator basis) is \cite{badasc}

\bear
{\cal L}_{\rm scal}^{(DH)} &=& 
\frac{e^2}{180 (4\pi)^2m^2} 
\biggl\lbrack
15(\xi - \frac{1}{6}) R F_\mn^2 
- 2 R_\mn F^{\mu\alpha}F^{\nu}{}_\alpha \non\\
&&\qquad\qquad
- R_{\mu\nu\alpha\beta}F^\mn F^\ab
+3(\nabla^{\alpha}F_{\alpha\mu})^2
\biggr\rbrack \, .
\non\\
\label{drumhathscal}
\ear
(the parameter $\xi$ refers to a non-minimal coupling of the scalar).
In \cite{davsch} two of the present authors presented the next order in the expansion of the 
effective Lagrangians obtained in \cite{badasc} in powers of the field strength, 
i.e. the terms of order $RF^4$ (there are no order $RF^3$ terms for parity reasons).

The purpose of the present letter is twofold. First, we will use the above effective actions
at the $RF^2$ level to compute the low energy limits of the one graviton - two photon amplitudes
with a scalar and spinor loop, and show that they relate to the four photon amplitudes in a KLT - like
way. Second, as a preparation for the study of the higher-point cases we will derive the
Ward identities for the one graviton - $N$ photon 1PI amplitudes in general. 

\section{Ward identities for the 1PI one graviton -- N photon amplitudes}
\label{sectionward}
\renewcommand{\theequation}{2.\arabic{equation}}
\setcounter{equation}{0}

We derive the relevant Ward identities, generalizing the discussion
in \cite{phograv1}.
There are two types of Ward identities,
those derived from gauge invariance and 
those that follow from general coordinate invariance.

\noindent
Gauge transformations are defined by
\bear
\delta_G A_\mu = \partial_\mu \lambda \ , \quad
\delta_G g_{\mu\nu} = 0
\ear
with an arbitrary local parameter $\lambda $.
Then gauge invariance of the effective action
\bear
\delta_G \Gamma[g,A] =0
\ear
implies that 
\bear
\nabla_\mu \bigg (
{1\over \sqrt g }{\delta \Gamma\over \delta A_\mu } \bigg )=0  \ .
\ear
Similarly, infinitesimal reparametrizations are given by 
\bear
\delta_R A_\mu = \xi^\nu \partial_\nu A_\mu  + \partial_\mu \xi^\nu A_\nu
 \ , \quad \quad
\delta_R g_{\mu\nu} = \nabla_\mu \xi_\nu  + \nabla_\nu \xi_\mu
\ear
with arbitrary local parameters $\xi^\mu $. The 
invariance of the effective action 
\bear
\delta_R \Gamma[g,A] =0
\ear
now  implies
\bear
\nabla_\mu \bigg (
{2\over \sqrt g }{\delta \Gamma \over \delta g_{\mu\nu }}
+{1\over \sqrt g }{\delta \Gamma \over \delta A_{\mu}}
A^\nu \bigg )
- {1\over \sqrt g }
{\delta \Gamma \over \delta A_{\mu}} \nabla^\nu A_{\mu}
=0  \ .
\ear
The Ward identities thus obtained can be combined and written more 
conveniently using standard tensor calculus as follows
\bear
&& \partial_\mu  {\delta \Gamma\over \delta A_\mu } =0 \, ,
\label{idwardgauge}\\
&& 2 \partial_\mu
{\delta \Gamma \over \delta g_{\mu\nu}}
+{\delta \Gamma \over \delta A_{\mu}} \partial_\mu  A^{\nu}
+ \Gamma_{\mu\lambda}^\nu \bigg ( 2
{\delta \Gamma \over \delta g_{\mu\lambda }}
+ { \delta \Gamma \over \delta A_\mu } A^\lambda \bigg )
- {\delta \Gamma \over \delta A_{\mu}} \nabla^\nu A_\mu=0 \ .
\nonumber\\
\label{idwardgravstandard}
\ear
We remark that, alternatively,
the Ward identities from gauge invariance  can be used to simplify
the Ward identities from reparametrizations.
In fact, an infinitesimal reparametrization can be written as 

\begin{eqnarray}
&& \delta_R A_\mu = \xi^\nu \partial_\nu A_\mu  + \partial_\mu \xi^\nu A_\nu =
\partial_\mu (\xi^\nu A_\nu) + \xi^\nu F_{\nu\mu}  \\
&&
\delta_R g_{\mu\nu} = \nabla_\mu \xi_\nu  + \nabla_\nu \xi_\mu
\end{eqnarray}
where the first term in the last rule for $A_\mu$ can be interpreted as a gauge transformation.
The invariance of the effective action $\delta_R \Gamma[g,A] =0 $
now  implies (making use of $\delta_G \Gamma[g,A] =0 $ as well)
\begin{eqnarray}
\nabla_\mu \bigg (
\frac{2}{\sqrt g}{\delta \Gamma \over \delta g_{\mu\nu }}\bigg )
+\frac{1}{\sqrt g} {\delta \Gamma \over \delta A_{\mu}} F_\mu{}^\nu
=0  
\end{eqnarray}
i.e.
\begin{eqnarray}
2 \partial_\mu
{\delta \Gamma \over \delta g_{\mu\nu}}
+ 2 \Gamma_{\mu\lambda}^\nu 
{\delta \Gamma \over \delta g_{\mu\lambda }}
+ {\delta \Gamma \over \delta A_{\mu}} F_\mu{}^\nu=0 
\nonumber\\
\label{idwardgravsimp}
\end{eqnarray}
which of course is equivalent to (\ref{idwardgravstandard}).

Now we consider the special case of the correlation function 
of one graviton and $N$ photons in flat space.

\bear
\Gamma^{\mu\nu,\alpha_1\ldots \alpha_N}_{(x_0,x_1,\ldots ,x_N)}
\equiv  {\delta^{N+1}  \Gamma\over \delta g_{\mu\nu}(x_0) \delta A_{\alpha_1}(x_1)\ldots \delta A_{\alpha_N}(x_N)}
\Biggr |_{g_{\mu\nu} = \eta_{\mu\nu} , \,  A_{\alpha_1} = A_{\alpha_2}= \ldots = A_{\alpha_N} = 0 }\ .
\label{Auno}
\ear
Taking appropriate functional derivatives on the general Ward identities
(\ref{idwardgauge}),(\ref{idwardgravsimp}) we obtain the gauge Ward identities
\bear
&& \partial_{\alpha_i}^{(x_i)} \Gamma^{\mu\nu,\alpha_1\ldots \alpha_N}_{(x_0,x_1,\ldots ,x_N)}  = 0\, , \qquad
i = 1,\ldots , N, 
\label{wardgauge} 
\ear
and the gravitational Ward identities

\bear
&& \sum_{i=1}^N
{\delta^N  \Gamma\over \delta A_\mu(x_0) \delta A_{\alpha_1}(x_1)\ldots \widehat{\delta A_{\alpha_i}}(x_i)\ldots \delta A_{\alpha_N}(x_N)} \Biggr |
\bigl(\delta^{\alpha_i}_\mu \partial^\nu_{(x_i)} - \eta^{\alpha_i\nu} \partial_\mu^{(x_i)} \bigr) \delta^D(x_0-x_i) 
\nonumber\\
&& + 2 \partial_\mu^{(x_0)}\Gamma^{\mu\nu,\alpha_1\ldots \alpha_N}_{(x_0,x_1,\ldots ,x_N)}
 = 0  \ , \label{wardgrav}
 \ear
where the ``hat'' means omission.

\noindent
Fourier transforming the identities (\ref{wardgauge}),(\ref{wardgrav}) to momentum space 
\bear
\int dx_0dx_1 .. dx_N\, e^{ik_0x_0 +.. +ik_Nx_N}\, \Gamma_{(x_0,.., x_N)} 
= (2\pi)^D \delta(k_0+..+k_N)\Gamma [k_0,.., k_N]
\nonumber\nonumber\\
\label{fourier}
\ear
they turn into

\bear
&& k_{i\alpha_i} \Gamma^{\mu\nu,\alpha_1\ldots \alpha_N}[k_0,\ldots,k_N] = 0\, ,\qquad i = 1,\ldots, N,
\label{wardgaugek} \\
&& 2 k_{0\mu} \Gamma^{\mu\nu,\alpha_1\ldots \alpha_N}[k_0,\ldots,k_N]
+ \sum_{i=1}^N \Gamma^{\mu\alpha_1\ldots \widehat{\alpha_i}\ldots \alpha_N}
[k_0+k_i,k_1,\ldots,\widehat{k_i},\ldots,k_N]
 (\delta^{\alpha_i}_\mu k_i^\nu -\eta^{\alpha_i\nu} k_{i\mu} )
\nonumber\\
&& = 0 \ .
\label{wardgravk}
\ear
Thus the gauge Ward identity is transversal as in QED, while the gravitational Ward identity  
relates the one graviton -- N photon amplitude to the pure $N$ photon amplitudes.

\section{The graviton-photon-photon amplitude}
\label{threepointonshell}
\renewcommand{\theequation}{3.\arabic{equation}}
\setcounter{equation}{0}

\begin{figure}[ht]
  \centering
    \includegraphics[width=0.3
      \textwidth]{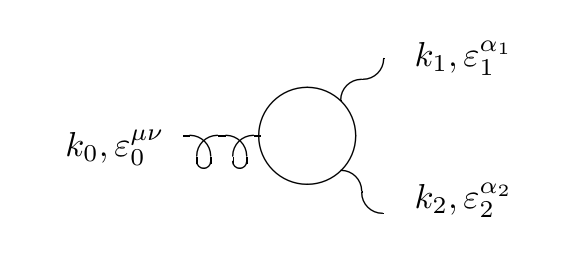}
        \caption{Graviton -- photon -- photon diagram} \label{gff}
\end{figure}

We proceed to the study of the on-shell graviton-photon-photon
amplitude induced by a scalar or spinor loop (see figure \ref{gff}).

First, let us remark that this amplitude does not exist at tree level for the fully on-shell
case. The covariantized Maxwell term in the action of Einstein-Maxwell theory

\begin{eqnarray} 
S[g,A] =    
\int d^D x\ \sqrt{g}\, \bigg (  
{1\over \kappa^2 } R - {1\over 4}F_{\mu\nu}F^{\mu\nu}  
\bigg )   
\label{EM}  
\end{eqnarray}
contains a graviton-photon-photon vertex, and this vertex with one photon leg off-shell is responsible for
the well-known process of photon-graviton conversion in an electromagnetic field \cite{gertsenshtein,rafsto,phograv1,phograv2,ahjari}.
However, it vanishes with all legs on-shell (let us also remark that, fully off-shell, this vertex provides already
an example for the non-trivialness of the gravitational Ward identity (\ref{wardgravk})). 

The low-energy limit of the one-loop amplitudes is readily obtained from the Drummond-Hathrell form
of the effective Lagrangians, (\ref{drumhath}) resp. (\ref{drumhathscal}).
In the on-shell case, the $R$, $R_{\mu\nu}$ and $ (\nabla^\alpha F_{\alpha \mu})^2$ terms all vanish.
The remaining term can, after the usual procedure of taking the Fourier transform and then 
truncating to lowest order in momenta \cite{phograv1}, be written as

\begin{eqnarray}\label{ampDH}
R_{\alpha \beta \mu \nu}\,F^{\alpha \beta}\,F^{\mu \nu}
&=&\varepsilon_{0 \mu \nu}\,\Gamma^{\mu \nu ; \alpha \beta}\,
\varepsilon_{1\alpha}\,\varepsilon_{2 \beta}
\nonumber\\
\Gamma^{\mu \nu ; \alpha \beta}&=&4\Big[ -k_1^{(\mu}\,k_2^{\nu)}\,k_0^\alpha\,k_0^\beta
+\eta^{\alpha (\mu}\,k_2^{\nu)}\,k_0^\beta\,k_0\cdot k_1\nonumber\\&&
+\eta^{\beta (\mu}\,k_1^{\nu)} \,k_0^\alpha \,k_0 \cdot k_2
-\eta^{\alpha (\mu}\,\eta^{\nu) \beta}\,k_0 \cdot k_1\,k_0\cdot k_2\Big]
\label{contRFF}
\end{eqnarray}
At the $N=2$ level the purely photonic terms in the gravitational Ward identity~(\ref{wardgravk})
vanish on-shell. Thus this identity holds in its usual form
$k_{0\mu}\,\Gamma^{\mu\nu;\alpha \beta}= 0$, as can be checked with~(\ref{contRFF}).

Proceeding to the helicity decomposition of the amplitude, using a factorized
graviton polarization tensor as usual,

\begin{eqnarray}
\varepsilon_{0\mu\nu}^{++}(k_0) &=& \varepsilon_{0\mu}^+(k_0)\varepsilon_{0\nu}^+(k_0)\,,\nonumber \\
\varepsilon_{0\mu\nu}^{--}(k_0) &=& \varepsilon_{0\mu}^-(k_0)\varepsilon_{0\nu}^-(k_0)\,, \nonumber\\
\label{polgrav}
\end{eqnarray}
together with  a judicious choice
of the reference momenta one can easily show that, of the six components of this amplitude,
only the MHV ones are nonvanishing
\footnote{It should be mentioned that even for these components the right hand sides will vanish
after taking into account that for a massless three-point amplitude energy-momentum conservation
forces collinearity of the three momenta. However, this is a low-point kinematic accident and not
relevant for our structural investigation.}
:

\begin{eqnarray}
\varepsilon^{++}_{0 \mu \nu}\,\Gamma^{\mu\nu ; \alpha \beta}\,\varepsilon^+_{1 \alpha}\,\varepsilon^+_{2 \beta}
&=& - [0 1]^2[ 0 2]^2 \nonumber\\
\varepsilon^{- -}_{0 \mu\nu}\,\Gamma^{\mu\nu ; \alpha \beta}\,\varepsilon^-_{1 \alpha}\,\varepsilon^-_{2 \beta}
&=&- \langle 0 1\rangle^2 \langle 0 2\rangle^2 \nonumber\\
\label{3pMHV}
\end{eqnarray}
Including the prefactors in (\ref{drumhath}),(\ref{drumhathscal}), and restoring the
coupling constants, we obtain the final result,

\begin{eqnarray}
A_{\rm spin}^{(++;++)}&=&-\frac{\kappa\,e^2}{90 (4\pi)^2m^2}\,[01]^2\,[02]^2\nonumber\\ %\,\langle k^+_0|k^-_1 \rangle^2\,\langle k^+_0|k^-_2 \rangle^2\,,\\
A_{\rm spin}^{(- - ; - - )}&=&-\frac{\kappa\,e^2}{90 (4\pi)^2m^2}\,\langle 01 \rangle^2\,\langle 02 \rangle^2% k^-_0|k^+_1 \rangle^2\,\langle k^-_0|k^+_2 \rangle^2\,,\non
\nonumber\\ 
\label{gppcomp}
\end{eqnarray}
Here the first upper index pair refers to the graviton polarization, and $\kappa$ is the gravitational coupling constant.
Moreover, those components fulfill the MHV relation (\ref{relscalspin}),

\begin{eqnarray}
A_{\rm spin}^{(++;++)}&=&(-2)\,A_{\rm scal}^{(++;++)}\nonumber\\
A_{\rm spin}^{(- - ; - - )}&=&(-2)\,A_{\rm scal}^{(- - ; - - )} \nonumber\\
\label{gppMHV}
\end{eqnarray}
Also, these graviton-photon-photon amplitudes relate to the (low energy) four photon amplitudes in the following way:
From (\ref{EHamp}), (\ref{defchi}) the only non-vanishing components of those are

\begin{eqnarray}
A^{++++}[k_1,k_2,k_3,k_4] &\sim& [12]^2[34]^2 +  [13]^2[24]^2 +  [14]^2[23]^2 \nonumber\\
A^{++--}[k_1,k_2,k_3,k_4] &\sim& [12]^2 \langle34\rangle^2 \nonumber\\
A^{----}[k_1,k_2,k_3,k_4] &\sim&
\langle12\rangle^2\langle34\rangle^2+\langle13\rangle^2\langle24\rangle^2+ \langle14\rangle^2\langle23\rangle^2 \nonumber\\
\label{4photcomp}
\end{eqnarray}
Replacing  $k_1 \to k_0, k_2 \to k_0$ in the 4 photon amplitudes, the middle one of these three components becomes zero,
and the remaining ones become proportional to the corresponding components of (\ref{gppcomp}):

\begin{eqnarray}
A^{++++}[k_0,k_0,k_3,k_4] \sim& 2[03]^2[04]^2 & \sim A^{++;++}[k_0,k_3,k_4] \nonumber\\
A^{----}[k_0,k_0,k_3,k_4] \sim& 2\langle03\rangle^2\langle04\rangle^2& \sim 
A^{--;--}[k_0,k_3,k_4] \nonumber\\
\label{gravtophot}
\end{eqnarray}
Thus in all cases one finds the same proportionality, namely

\begin{eqnarray}
A_{\rm scal,spin}^{\pm\pm;\pm\pm}[k_0,k_1,k_2] = \frac{1}{12}\frac{\kappa}{e^2}m^2 A_{\rm scal,spin}^{\pm\pm\pm\pm}[k_0,k_0,k_1,k_2] 
\label{universal}
\end{eqnarray}
Effectively two photons have merged to form a graviton, clearly a result in the spirit of the new  KLT-like  relations.
In \cite{stieberger} the same (except for the proportionality constant) relation was found for the graviton-photon-photon
amplitude in superstring theory at the tree level. 

We have derived our results using the low energy effective action, but it is straightforward to reproduce them
using worldline methods, where the gravitational amplitude $A^{\pm\pm;\pm\pm}$  is obtained by computing,
in the low energy limit, a correlation function of the corresponding vertex operators of the form
\begin{equation}
\int_0^\infty \frac{dT}{T} \frac{{\rm e}^{-m^2 T}}{(4\pi T)^2} \left ( -\frac{\kappa}{4 T} \right ) (-i e)^2
\biggl \langle  V_{grav}(k_0) V_{ph}(k_1)V_{ph}(k_2) \biggr \rangle
\end{equation}
Indeed, we hope to use such methods to investigate higher point amplitudes.

\section{Conclusions}
\label{conclusions}
\renewcommand{\theequation}{4.\arabic{equation}}
\setcounter{equation}{0}
We have shown here that, in the low energy limit, the one-loop graviton-photon-photon 
amplitudes in Einstein-Maxwell theory coupled to scalars or spinors relate to a coincidence
limit of the QED four photon amplitudes. This provides a new example of a KLT-like factorization
in field theory at the loop level, and agrees with an identity
found in \cite{stieberger} at the tree level in superstring theory.
It also raises the possibility that, at least in the low energy limit, the $M$ graviton -- $N$  photon amplitudes 
may be derivable from the $N+2M$ photon amplitudes. However, it must be emphasized that the three-point amplitude
is rather special in this context due to the absence of one-particle reducible contributions.
The inhomogeneity of the gravitational Ward identity (\ref{wardgravk}) leads one to expect that,
starting from the one graviton -- four photon level, the 1PI one graviton - $N$ photon amplitudes will not be transversal
in the graviton indices,  so that a relation with the purely photonic amplitudes can exist only for the full
amplitudes. The calculation of the one graviton -- four photon amplitude is in progress.

\medskip

\noindent
{\bf Acknowledgements:}
F.B. and C.S. thank S. Theisen and the AEI Potsdam for hospitality
during part of this work.
The work of F.B. was supported in part by the MIUR-PRIN contract 2009-KHZKRX.
The work of O.C. was
partly funded by SEP-PROMEP/103.5/11/6653. C.S. was supported by
CONACYT grant CB 2008 101353.

%%%%%%%%%%%%%%%%%%%%%%%%%%%%%%%%%%%%%%%%%%%%%%%%%%%%%%%%%%%%%%%%%%%%%%%%%%%%%%%

\end{document}